# Group performance is maximized by hierarchical competence distribution

Anna Zafeiris[1] & Tamás Vicsek*[1,2]

[1]*Department of Biological Physics, Eötvös University, Pázmány Péter sétány 1A, H-1117, Budapest, Hungary.* [2]*Statistical and Biological Physics Research Group of HAS, Pázmány Péter sétány 1A, H-1117, Budapest, Hungary.*
*vicsek@hal.elte.hu

**Groups of people or even robots often face problems they need to solve together. Examples include collectively searching for resources, choosing when and where to invest time and effort, and many more. Although a hierarchical ordering of the relevance of the group members' inputs during collective decision making is abundant, a quantitative demonstration of its origin and advantages using a generic approach has not been described yet. Here we introduce a family of models based on the most general features of group decision making to show that the optimal distribution of competences is a highly skewed function with a structured fat tail. Our results have been obtained by optimizing the groups' compositions through identifying the best performing distributions for both the competences and for the members' flexibilities/pliancies. Potential applications include choosing the best composition for a group intended to solve a given task.**

Group life involves a continuous series of collective decision making events related to a large selection of tasks[1-3], such as searching for food[4], navigating towards a distant target[5-8] or deciding when and where to go[7,9]. The members of a group typically contribute to finding the best solution with varying degrees of input, because of the engineered or naturally occurring differences in their capabilities of possessing information [1,10-16]. Recent theoretical interest focused on two possible mechanisms of group decision making[13, 17-19] based on the influence of the members originating from, e.g., their level of dominance, physiological state or pertinent information and/or navigational competence[7]. In the "democratic" or egalitarian version the members contribute to the final decision to about the same degree[20], while in a "despotic" situation one or a few individuals play the role of leaders and determine the final outcome of the decision process[21, 22]. It has been observed experimentally that the latter kind of influence allocation may increase the efficiency of a group[6, 22]. Up to very recently[5] when addressing the role of leadership in animal groups quantitatively, the simplest case has been considered, with one or more "informed" individuals (e.g., pre-trained fish or birds), while the rest of the members played the role of followers. Due to the sensitivity and the effectiveness of group decision making this simple "2 level hierarchy" has already led to interesting findings both using a modelling[17] and experimental[22] approaches. All these works were aimed at finding/interpreting the evolutionary stable (optimal) solution based on individual selection.

At the same time, recent experimental observations involving some sophisticated animal groups such as pigeons or primates point towards the possibility of significantly more complex internal organization principles[5, 23, 24]. In socially highly organized groups beyond a given size (dozens or so) the roles related to leadership do not seem to be simply binary, but several levels of hierarchy can be identified. This is how groups of apes, organizations, or even a group of pigeons behave. While in prior works two-level hierarchies (with two, well distinguished kinds of group members: the leaders and followers) have been considered, here we demonstrate that a multiple-level hierarchy is likely to be more optimal in some cases. We explain this result with the spreading (mixing) of the information between the individuals, which is much more efficient in a system of multi-level hierarchical interactions than in a two-level (or "bimodal").

Motivated by the above fundamental considerations, we have decided to address the problem of identifying the *optimal distribution of the competence and pliancy values of the individuals within groups* (exceeding the size of a few dozen) *that are faced with a problem* different from just staying together. The members do not have the knowledge of the competence of the others, they do not distinguish each other, and they interact according to an underlying network. In order to reveal these optimal distributions, we measure the 'quality' of the solution provided by the group and correlate it with the competence levels of the members. In our interpretation, competence corresponds to the level of the ability of an agent to facilitate the solving of a problem and pli-ancy refers to the willingness of an individual to follow others (mostly neighbours). In our case optimal performance is associated with finding the best solution (i.e., gaining the largest amount of benefit) using the smallest amount of cost. Competence appears as a cost, because it requires learning, experience or knowledge requiring investments.

Recently, there has been a growing interest in models with similar assumptions, focusing on the optimal strategies and characteristics adopted by self-interested individuals. These related fields include the topic of target seeking, coordination and the so called "producer-scrounge" game. We shall overview these results in the Discussion section.

In order to relate our work aimed at a more abstract set of problem solving situations than the one associated with the particular (and much studied, interesting) topic of target seeking[5,6,8,21,22,25] we have carried out numerical experiments on the latter problem as well. However, we would like to stress here as well that the prime intention of our study has been to indentify the optimal competence and pliancy distributions *over a wider range of problems and communication networks*. In other words, target seeking is only a special case in our study, while the main goal is to determine the existence and nature of a general/universal competence versus pliancy distribution which would ensure optimal or near-optimal problem solving behaviour in various kinds of groups.

In the present paper we study order hierarchy (hierarchy from now on) being equivalent to an ordering induced by the values of a variable (in our case competence) defined on some set of elements. We are aiming at determining the best distribution of competences in a group under the condition that the total resources used (sum of competences) for achieving a given goal should be as small as possible. Here we introduce a family of models based on the most general features of group decision making to show that – from an approach based on first principles only – the optimal distribution of competences is a highly skewed function with a structured fat tail. Thus, the amount of resources (information, cost, knowledge) needed for finding a good solution by, e.g., a group of people or robots is minimized when the group's competence levels are hierarchically ordered. We show that such a distribution leads to performances considerably exceeding those obtained for other common distributions. Our finding emerges from the interaction dynamics within the collective. It is highly robust, being nearly independent of the number of group members, the kinds of problems to be solved and the structure of the underlying network of interactions. A counterintuitive, but reproducible feature of our findings is a hump in the tail of the distribution function. These results were obtained by optimizing the group behaviour of our models by identifying the best performing distributions for both the competences and for the members' flexibilities/pliancies (willingness to comply with other group members)[26].

**Results**

   **Basic features and procedures of the generic group decision making models we study**

      Next we need to summarize the basic features of the way a group approaches its best answer during a collective decision making process. We consider decisions which emerge from the instantaneous estimates of the group members concerning the best choice to proceed or, alternatively, about the final solution. One of our main observations/statements is that most of the tasks to be completed by collective decision making can be reduced to this "estimation" paradigm. We consider the following general situation: finding the best solution happens in rounds of interactions during which

      - each individual makes an estimation of the best solution based on its competence (ranging from small to very good), and from the behaviours of its neighbours (neighbours being represented by nodes of various networks).
      - the actual choice of the members also depends on their varying flexibilities (pliancies, i.e., the level to which they are willing to adopt the choices of their neighbours)
      - a collective "guess" about the true solution is made.

The performance of a group is measured after each run / trial. The best distribution of a group is approached by varying the distribution of competences and pliancies making use of a genetic algorithm[26]. The process of problem solving is stopped after some simple criteria are satisfied, e.g., the guesses converge, a given number of time steps is reached or the guess achieved a pre-defined accuracy. The optimal distribution is then associated with the average distribution of the competence and pliancy values appearing in the 500 best performing (most optimal) groups.

Thus, we define several (four) Group Performance Maximization models (GPM models or GPMMs). In these models each group has to solve a model-dependent problem (for the flowcharts see Supplementary Figs. S3-S7). Because of the simplicity of our GPMMs, many real-life tasks can be mapped on each of them. The quality of the groups' performance, *Pe*, is quantifiable and characterized by a parameter with values in the *[0, 1]* interval. Higher values correspond to better performance. The contribution of the $i$th group member to finding the best solution depends on its competence level $Co_i$. $Co_i$ also takes values from the *[0, 1]* interval. Each model consists of iterative steps. The behaviour $Be_i^{(t+1)}$ of agent (member) $i$ at time step $t+1$, depends both on its own estimation $f(Co_i)$ regarding the correct solution, and on the (observable) average behaviour of its neighbours $j(\epsilon R)$ in the previous step $t$, $<Be_j^t>_{j\epsilon R}$:

$$Be_i^{(t+1)} = (1-\lambda_i)f(Co_i) \oplus \lambda_i <Be_j^t>_{j\epsilon R}, \qquad (1)$$

$\oplus$ denotes "behaviour-dependent summation", where "behaviour" refers to various actions, such as estimating a value, casting a vote or turning into a direction, etc. The set of weight parameters $\lambda_i$ takes values on the *[0, 1]* interval and defines the pliancy distribution. Some kind of noise, explicitly or implicitly, was incorporated into all models (for details of the models see the Supplementary Information).

**Models with pre-defined static communication networks**

We first focus on models in which the interactions are defined by static networks. In order to elicit the possible effects of the communication structure we have studied these models using several network types, such as small-world, hierarchical[27], Erdős-Rényi and a real-life social network describing the friendship relations in a school[28]. Rather counterintuitively[29] - our findings are quite independent of the type of the networks used as we demonstrate displaying the results for all graph types (Fig. 2 and Supplementary Fig. S1).

To see how the nature of a problem affects the optimal competence distribution we have calculated the performances for the following GPM models: (i) the Voting GPMM, which was designed to be as simple as possible, (ii) the Sequence guessing

GPMM, designed to be still simple, but widely applicable, and finally (iii) the Direction finding GPMM, corresponding to a less abstract situation.

In our Voting GPMM – having some analogy with the widely used Ising model – , the group has to find the correct answer choosing from two options (yes/no, -1/1, etc). This minimal model consists of two steps only (see Supplementary Fig. S4). In the first one all agents make a guess being correct in proportion of their competences. Once this is done, the actors count the guesses of their neighbours and based on this they cast a vote. This second step is realised according to Eq. 1. Figure 1c shows the optimal competence distribution for all $\lambda_i=1$, that is, when the choices of the neighbours determine the vote of an individual. This distribution ensures the highest rate of voting correctly, and thus, the highest group performance as well.

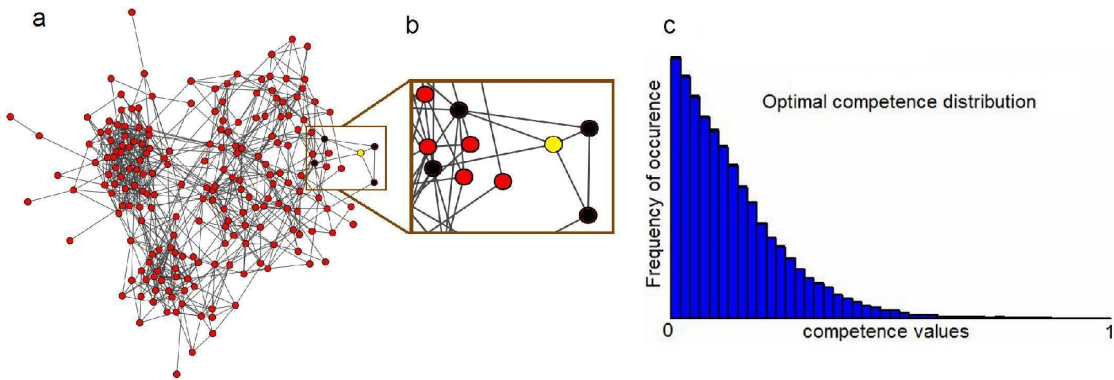

**Figure 1**. **The Voting GPM model. a,** The "Friendship graph", a real-world social network reflecting the amity relations in a high school among 204 students[28]. **b,** An enlarged portion of the network showing the influential relations from the viewpoint of the node coloured yellow. **c,** The optimal competence distribution for the Voting GPM model: a highly skewed function with a fat tail.

The second GPM model was designed to solve the most general problem we could think of in our context, i.e., the estimation of a series of numbers. We argue that most of the simple tasks can be mapped onto this problem, including estimating a direction, or finding a location (given by direction and distance) or even estimating the distribution of incomes from various sources, this is why similar problems have been widely studied by economists as well[30]. In our model a sequence of real numbers (between 0 and 1) had to be estimated iteratively (Supplementary Fig. S6). In each step each actor modified its actual guess for each element of the number sequence. This modification depended on two circumstances: its own estimation and the corresponding average guesses of its neighbours. Eq. 1 describes this process, if $f(Co)$ is interpreted as a 'value guessing function' returning more precise results for higher $Co$ (competence) input values and '$Be$' is interpreted as the act of adopting a value. The length of the number sequence corresponding to the results displayed in Fig. 2 was 10, but again, the number of numbers to be guessed did not have a significant effect on the outcome.

Finally, the third GPM model was designed to address a less abstract situation in which the group had to find out a pre-defined direction (Supplementary Fig. S5). If the $f(Co)$ function returns a direction estimation and $Be$ is a direction (vector of unit length), then Eq. 1 describes the process.

The optimal competence distributions for all three models and all four network types are summarized in Figure 2 and Supplementary Figure S1. We have obtained these results by using a genetic algorithm[26] in which the fitness function $F$ was defined as

$$F = Pe - K<Co>, \tag{2}$$

where $K$ is a parameter reflecting the "cost of learning" and $<Co>$ is the average competence level of the group. As the figures demonstrate, the competence values form a hierarchically ordered distribution in all cases, with progressively fewer members having high competence values except a specific deviation from this rule.

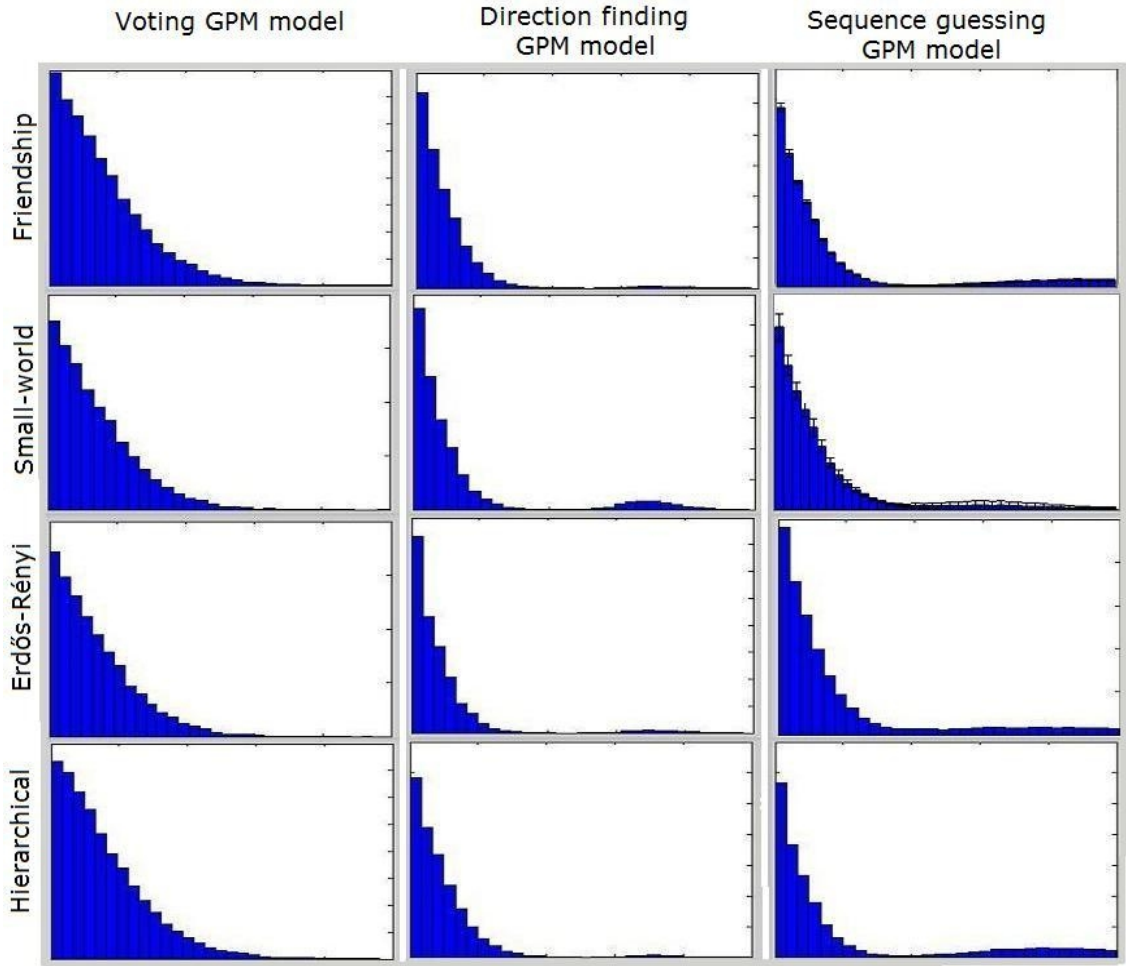

**Figure 2**. **The optimal competence distributions for the three models and 4 kinds of networks.** These optimal distributions are hierarchically ordered, highly skewed functions often with a structured tail. The size of the "Friendship" network is $N$=204, while the other graphs contain $N$=200 nodes, and $K$ was set to 2. We have studied networks with various sizes, ranging from $N$=10 to 200 finding no significant change in the shape of the distributions. In the last column (belonging to the Sequence guessing GPMM) we have marked the error bars for the Friendship and small world networks (the error bars for the rest of the plots fall into the same range).

Here we investigate problems being abstract to a different degree. In the case of the less abstract ones (like the flocking game), the "optimal solution" maximally satisfies a combination of several "*intuitively favourable*" conditions, such as (a) the average speed with which the target is reached, (b) the ratio of the flock that does not get lost, and (c) level of cohesion of the group (information being able to spread over the whole flock. In the case of those models where the problems themselves are more abstract mappings of real-life problems (such as the Sequence guessing model), the formulation of the "optimal solution" becomes inevitably more abstract as well. However, in such cases, the *improvement rate* of the average estimation of the group is the quantity we optimize since it is a good measure of the *efficiency* with which the group approaches the true solution.

Our calculations show that in the case of the Direction finding GPMM and Sequence guessing GPMM, the fat tails are structured, having a smooth "hump". In order to show the extent to which the optimal distributions improve the group performance we have calculated *Pe* for a few known distributions of competences as well. Figure *3f* shows the results for the real-life social network, "Friendship", and for the most realistic GPM model, the Direction finding GPMM. The average competence level is identical in all cases. We conclude that the simultaneous choice of both the competence and the pliancy distributions are essential, and the optimal choice results in a strong improvement of the efficiency. Another observation is that – somewhat counter-intuitively – the particular structure of the underlying network of interactions does not have a relevant effect on our basic finding.

Importantly, the results depicted on Figure 2 are achieved by letting the pliancy values – denoted by λ in Eq. 1 – evolve simultaneously and independently from the competence values. Figure 3 reveals the relation between the competence and the pliancy values within an optimized group. The first conspicuous result is that – in analogy with the competence values – the pliancy values also form a highly skewed distribution. However, in this case actors with high pliancy values form the majority (Fig. 3e). Figure 3a depicts how the average pliancy value (marked with thin pink solid line) steadily grows from generation to generation.

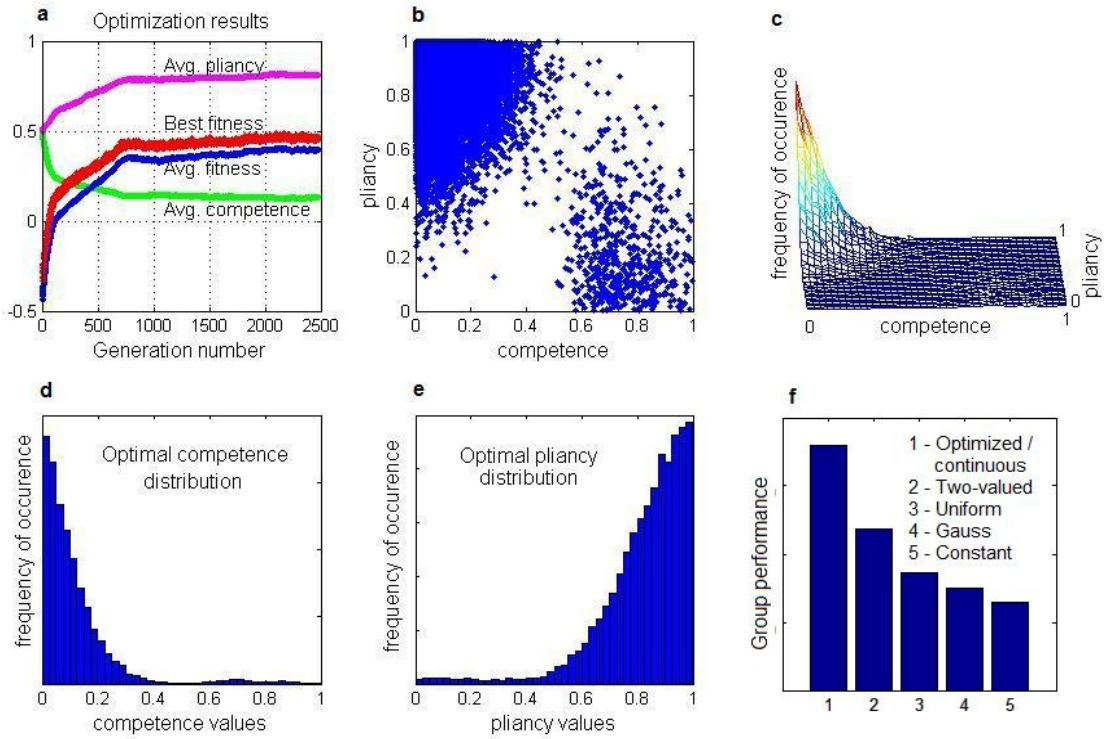

**Figure 3**. **Detailed results for the Direction finding GPMM on the Friendship graph.**
$K$=2. **a,** The first 2500 generations of the optimization algorithm, **b,** competence-pliancy
values depicted on a unit square. Each dot represents an agent. **c,** the same as **b,** but here
axis $z$ depicts the density of points. The uninformed individuals (low competence, high
pliancy) strongly outnumber the ones who are competent. **d,** Optimal competence distribu-
tion. **e,** Optimal pliancy distribution. **f,** comparison of the group efficiencies $Pe$ after 20
steps of iteration for the one we find and a selection of commonly assumed distributions.
From left to right: Optimized/continuous, two valued (allowed competence values were
0.1 and 0.9), uniform, Gaussian, and constant. In order to demonstrate the effect of the dis-
tribution of the competence values more clearly, the pliancy values were set to be antagon-
istic for all the five cases (according to $\lambda_i=(1-Co_i)+\xi$ ).

Regarding the relationship between the competence and pliancy values, Figure 3b
and c grants a deep insight concerning their connection, and sheds light onto the origin
of the "hump" as well. The location of a point in Figure 3b is determined by the given
agent's competence ($x$ axis) and pliancy ($y$ axis) values. Two kinds of actors appear in
Fig. 3b and c: one kind clusters in the top left corner, corresponding to small compet-
ence and high pliancy values (these actors have "sheep mentality", and significantly
outnumber the rest of the agents), while the other kind has considerably higher compet-ence values mostly coupled with small pliancy characteristics. The hump – observable in most of the competence histograms – is due to the second kind of agents.

### Relation to the target seeking problem

In order to relate our findings to the much explored topic of target seeking, we have conducted simulations in which a group of agents, moving on a two dimensional surface, had to reach a pre-defined target in the shortest possible way.

Here the interaction among the group members change dynamically according to the actual distance among them: only those agents can exchange information with one another that are closer than a pre-defined distance called range of interaction, ROI (Fig. 4 and Supplementary Fig. S7). This distance is often associated with the range of vision among the individuals.

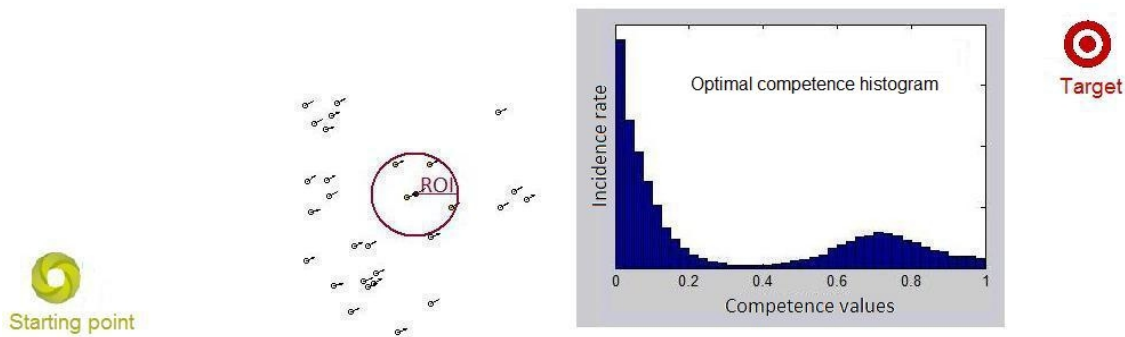

**Figure 4**. **The Flocking GPM model.** The distribution of the competence values is a highly skewed function in this case too, with a structured tale.

The pliancy values are set to be antagonistic to the competence values, according to: $\lambda_i = (1 - Co_i) + \xi$ for all agents. $\xi$ is random noise taking values uniformly in the (-0.1, 0.1) interval.  As it can be seen on Fig. 4, we find two groups as well (the one comprising the highly competent but 'anti-social' individuals, and the other containing the ignorant but pliant members[25, 31]), but in a much more smoothly distributed way.

**Continuous vs. bimodal competence distributions**

We believe that the reason behind the high group performance associated with a more continuous competence distribution (more continuous than that in the bimodal case) is due to a phenomenon that we call "information spreading or mixing", which can be summarized as: *Multi-level hierarchical interactions make the spreading (mixing) of the information between the individuals much more efficiently than in a "two-level" system.*

This interpretation is based on the following assumptions: (i) the individuals do not have knowledge of the level of competence of the others, (ii) the pliancy values change oppositely with the competence values (which is the general assumption in two-level systems), and finally, (iii) not all of the members interact with all of the other members, but according to an underlying network (which is again a natural assumption for groups beyond a given size). Given these observations, the bimodal competence distribution can often result in permanently or temporarily segregated groups maintaining different "opinions" or estimates of the true solution.

The reason behind this possibility of segregation is that uninformed individuals have a strong tendency to follow the others (since they have large pliancy values). Subgroups of the whole group thus can easily agree on a wrong estimate and they will maintain that until a better estimate "diffuses" to them from other groups having highly competent individuals. In the case of the homing flock model the segregation of groups can take place for a very long time, since in that case the network of interactions is such that groups moving in different directions lose contact completely.

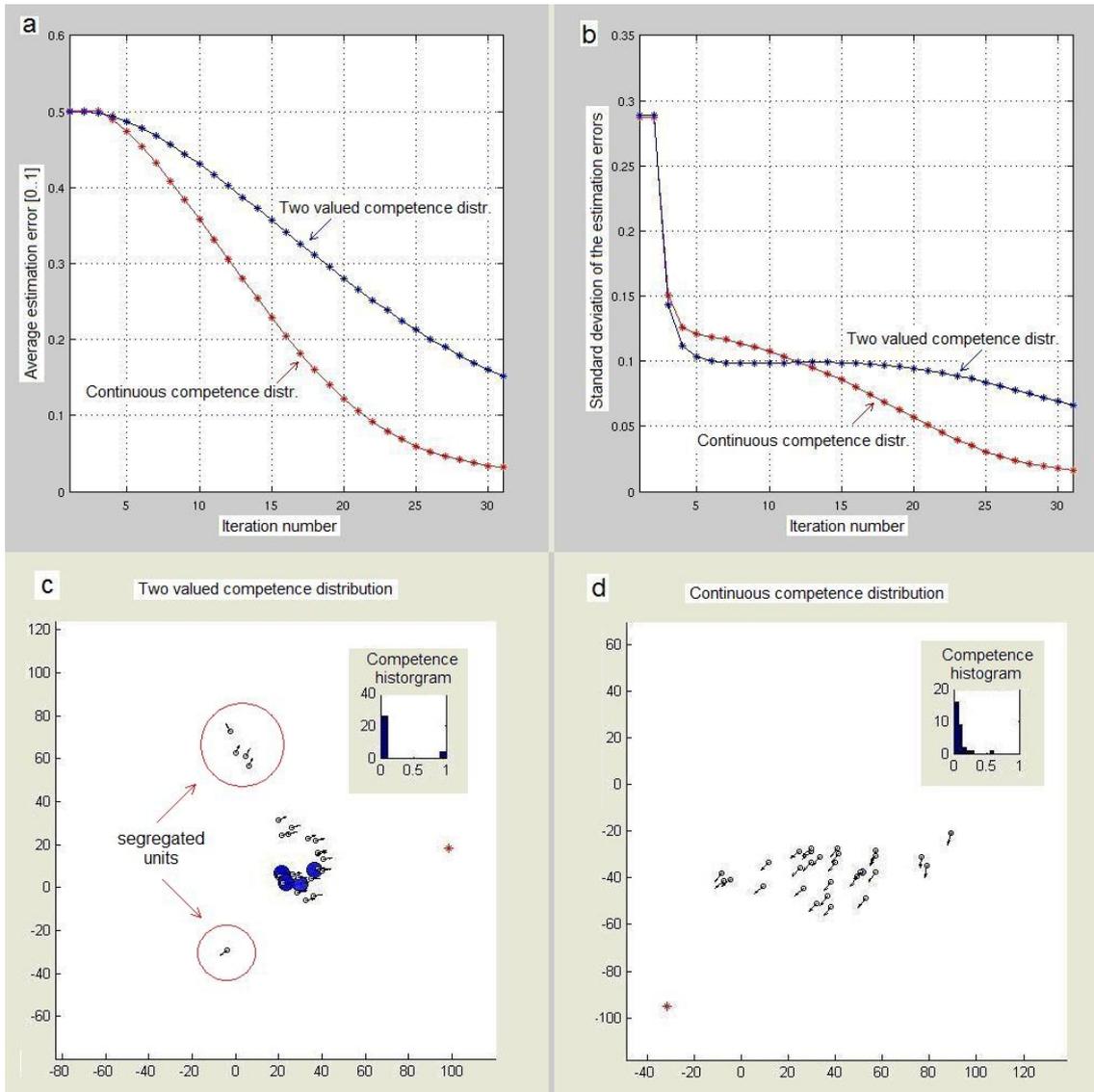

**Figure 5. A demonstration of the information mixing concept. a,** The average estimation errors as a function of the iteration number for the two-valued (blue curve) and for the continuous (red curve) competence distributions, for the sequence guessing model and for the Friendship graph. Each curve is the average of 100 test-runs. **b.** Although in the case of the two-valued competence distribution the standard deviation of the estimation errors drops more quickly initially, this fast decay stops after a few steps, while in the case of the continuous competence distribution we observe a steady decay, implying a more coherent and uniform change in the estimations. **c,** The lack of this unified change in the estimations might lead to the split-up of a flock in case the units are moving, **d,** while a smoother hierarchical decay of the competence distribution keeps the flock together.

**Discussion**

One of the possible collective decision making situations, the target seeking (or migration) problem has recently been addressed in detail.  In this case, the aim of the members is to reach the target in the shortest possible way. The competence level of each unit refers to the accuracy of the knowledge regarding the position of the target. These competence levels have been reported to be distributed strongly unevenly by several experimental[5,12] and theoretical[17] studies. Furthermore, members have different tendencies to follow others, which are typically assumed to be directly related to their competence levels[7, 25].

In this context, the optimal strategies adopted by migrating animals have been studied in a model in which individuals moved in a direction determined by the balance of two factors: (i) their preferred direction (which depended on their "gradient detection ability") and (ii) the direction of the other group members[25]. Both abilities (gradient detection and sociality) came at a cost. A strategy, adopted by an individual, was considered to be optimal if the corresponding fitness - defined by the migratory benefits minus the costs - was maximal. Two well-defined, coexisting strategies have been found, resulting in a collective migration expressing the characteristics of a fission-fusion process: an individual either invested in acquiring information about the best migratory direction (these individuals were much less prone to follow others) or alternatively, adopted a socially facilitated motion, that is, exploited the ones who invested in "learning". The appearance of a specialized group of leaders within a migrating flock has also been found by analytically solving the Kuramoto type situation (everyone interacts with everyone) of the target seeking problem[31]. These studies, as well as the ones reviewed hereafter, assume individual selection, that is, all the costs and benefits are associated with the individuals, and with them alone.

The problem of coordination[32] (in which the members of a group, although they prefer to stay and act together, differ in their preferred course of action) is in close connection with the more general question of the origin and emergence of dominance hierarchies, which has attracted a lot of attention for a long time[33-35], and for which various models had been proposed[36-38]. However, although related, the issue of dominance hierarchy is also different from our question of interest, because individuals, when facing a given problem, do not necessarily copy the behaviour of the dominant ones, but they tend to copy the acts of the most competent group members[5]. This follows from the fact that dominant individuals are not necessarily the most competent ones regarding all possible abilities simultaneously.

Among the related topics, we should mention the problem known as the "producer-scrounger" game as well[39-41]. Here the producers search for food (explore the environment on their own cost) while scroungers, not willing to pay the cost of exploring, just take the food the producers have found. The trait that sets the two problems apart is that in the "producer-scrounger" game the scroungers (who are the "followers" in the target seeking problem) do not merely follow the producers ("leaders"), but also take away benefits from them.

There are several novel features of the above results. Our simulations indicate the robustness of the one-sided nature of the optimal distribution of competences and pliancies in groups solving a variety of problems. Although this is not against intuition, our study is the first one which provides this result from a quantitative analysis. Another important new point is that we have optimized the performances of the *groups* as whole but gained information about the properties of their *members.* Our work provides a framework for treating a wide selection of phenomena including several recent observations, eg., it is related to the problem of a few well-informed individuals being able to lead a group of  individuals  efficiently[1,2,12,21], the observation made by company managers that a group of skilled workers do not perform better than a group of workers with diverse abilities[10,42] and the results of models optimizing the strategies of individuals performing a specific task as part of a collective. Our finding emerges from the interaction dynamics within the collective.

The results we present are not in contradiction with the findings of studies assuming individual selection, but rather complement them (see Supplementary Information, Fig. S8). The main feature of the competence distribution we obtained is highly robust, being nearly independent of the number of group members, the kinds of problems to be solved and the structure of the underlying network of interactions. Knowing the optimal distribution of competences in model systems provides a deeper insight into determining the best performing distribution of a group even if in many applied situations the ac-tual tasks and conditions finally lead to decisions only remotely resembling the theoretically best choice.

**Methods**

All four GPMMs were coded in Matlab. Except for the Flocking GPMM, the agents interacted along the edges of a graph. The results shown were obtained for four basic network-types: (i) a real-world network reflecting the friendship relations in certain American high-schools[28], referred to as "Frnd", (ii) a hierarchical network referred to as "Hier", the undirected version of a graph generated by the algorithm described in[27] with hierarchy parameter $h^{Hier}=0.8$, (iii) Erdős-Rényi graph with edge-probability parameter $p^{ER}=0.015$, and (iv) Small-world graph with *15%* of the edges being randomized. The node and edge numbers were: $N^{Frnd}=204$, $E^{Frnd}=1012$, $N^{Hier}=200$, $E^{Hier}=777$, $N^{ER}=200$, $E^{ER}{\sim}=310$, $N^{SW}=200$ and $E^{SW}{\sim}=400$, respectively.

All optimizations were carried out using a genetic algorithm (GA)[26]. More precisely, we have used a standard GA with the distinction of using real numbers instead of a binary string of *0*s and *1*s, defining a phenotype. Each phenotype consisted of an array of *2N* elements (taking values from the [0, 1] closed interval), where *N* was the size of the group we aimed to optimize ($N{\approx}200$, see above). The first *N* values defined the competence level of the group-members; the second *N* numbers determined their pliancy values. The fitness function was defined as $F=Pe-K{<}Co_i{>}$ in all the four models, however, the concrete definition of the group performance *Pe* differed from model to model. In the above equation, *K* is a parameter describing the "cost of learning" (typically 1 or 2), and $<Co_i>$ is the average competence level within the group. The actual definition of the *Pe* functions was much more dependent on the given models, but as a rule, all of them returned a value between 0 and 1, with higher values indicating better group performances. The fitness functions (and within the fitness functions the group performances *Pe*) used by the genetic algorithm for optimizing the various GPMMs were designed to be as diverse as possible in order to avoid artificial similarity by designing them in a way that they would measure different *aspects* of collective problem solving.

Accordingly, in the 'voting model' it measured a *proportion* of the group (the ratio voting correctly), while in the 'direction finding model' it was related to the *improvement* of the average direction estimation (expressed in radians but divided by $\pi$ in order to scale it into the [0, 1] interval). Somewhat similarly, in the 'sequence guessing model' *Pe* reflected the improvement of the estimations too, but this time expressed as the ratio of the original estimation error. And finally, in the 'flocking model', *Pe* was a complex function depending on three factors: (i) how fast the group could reach the target, (ii) whether the flock stayed "connected" or some of the members had to navigating alone, and (iii) how many units got lost. By formula,

$Pe_{Voting} = N^{correct}/N,$

where $N$ was the total number of units, and $N^{correct}$ denoted the units voting correctly.

$Pe_{DirectionFinding} = (AvgerageDirectionError_{t=0} - AvgerageDirectionError_{t=tMax})/\pi,$

where the average direction errors are interpreted in radian,

$Pe_{SequenceGuessing} = (InitialGroupError - FinalGroupError)/InitialGroupError,$

where *GroupError*s denote the average estimation error of the group, and finally

$Pe_{Flocking}$ is the most complex one, including a factor describing how straightly/quickly the group reached the target: $D/<s_i>_T$, an other one expressing the level by which the flock was navigating together (in contrast with all the units searching for the target alone): $(1-<s_i^{out}/s_i>_T)$, and a third factor depending on the number of lost units: *(1-$N_{Lost}$/N).*

$Pe_{Flocking}= D/<s_i>_T (1-<s_i^{out}/s_i>_T) (1-N_{Lost}/N),$

where $D$ is the distance between the starting point and the target, $s_i$ is the number of steps needed by unit $i$ to reach the target, $s_i^{out}$ is the number of steps in which unit $i$ was "alone" (in the sense that its ROI was empty), $N_{Lost}$ is the number of lost units, $N$ is the number of all units (that is, the flock size), $<...>$ is averaging over all the units, and finally $<...>_T$ is averaging over those elements who reached the target.

For the Sequence guessing (SG), Voting, Direction finding (DF) and the Flocking models the applied generation numbers ($GN$) and population sizes ($PS$) were $GN^{SG}=3000, PS^{SG}=700, GN^{Voting}=7000, PS^{Voting}=600, GN^{DF}=2500, PS^{DF}=500, GN^{Flocking}=1100, PS^{Flocking}=500$, respectively, ensuring the proper saturation. For the above defined parameters, the required processor times ($t_{Proc}$) for the various GPMMs differed considerably: $t_{Proc}^{Voting} \sim 160$ minutes , $t_{Proc}^{SG} \sim 27$ hours , $t_{Proc}^{DF} \sim 50$ days and finally $t_{Proc}^{Flocking} \sim 45$ days. The type of the network did not affect these values considerably.

Here we would like to note that optimizing a problem by using a genetic algorithm that is related with groups has no *ab ovo* relation to group selection at all (in terms of the evolutionary theory of life). Although the wording is similar, and some of the technical assumptions are analogous to what is used in evolutionary theory, our approach is not related to, or involves "group selection".

The detailed descriptions of the four GPMMs along with the corresponding flow charts and fitness functions can be found in the Supplementary Information.

**References**

1.    Whallon, R. Lovis, W. A. & Hitchcock, R. (eds.) *Information and its Role in Hunter-Gatherer Bands (Ideas, Debates and Perspectives)* (Cotsen Institute of Archaeology Press, 2011).

2     Conradt, L. & List, C. Group decisions in humans and animals: a survey. *Phil. Trans. R. Soc. B* **364**, 719–742 (2009).

3     Vicsek, T. & Zafeiris, A. Collective motion. *Phys. Rep*. **517**, 71-140 (2012).

4     Viswanathan, G. M., da Luz, M. G. E., Raposo, E. P. & Stanley, H. E. *The Physics of Foraging: An Introduction to Random Searches and Biological Encounters*. (Cambridge University Press, New York, 2011).

5     Nagy, M., Ákos, Z., Bíró, D. & Vicsek, T. Hierarchical group dynamics in pigeon flocks. *Nature*  **464**, 890-893 (2010).

6     Biro, D., Sumpter, D. J. T., Meade, J. & Guilford, T. From Compromise to Leadership in Pigeon Homing. *Curr. Biol.* **16,** 2123–2128 (2006).

7     Petit, O. & Bon, R. Decision-making process: The case of collective movements. *Behav. Process*. **84**, 635-647 (2010).

8     Faria, J. J., Codling, E. A., Dyer, J. R, G, Trillmich, F. & Krause, J. Navigation in human crowds; testing the many-wrongs principle. *Anim. Behav.* **78,** 587-591 (2009).

9     Seeley, T. D. *Honeybee Democracy*. (Princeton University Press, New Jersey, 2010).

10    Surowiecki, J. *The Wisdom of Crowds (*Anchor, New York, 2005).

11    Page, S. E. *Diversity and Complexity* Primers in Complex Systems (Princeton University Press, New Jersey, 2010).

12    King, A. J. & Cowlishaw, G. Leaders, followers and group decision-making. *Commun. Integr. Biol.* **2,** 1-4 (2009).

13    Van Vugt, M. The Nature in Leadership: Evolutionary, Biological, and Social Neuroscience Perspectives. In: Day, D. V. & Antonakis, J (eds): *The Nature of Leadership*. (SAGE Publications, Inc, Thousand Oaks, california, 2011)

14    Piyapong, C., Morrell, L. J, Croft, D. P, Dyer, J. R. G, Ioannou C. C & Krause J. A Cost of Leadership in Human Groups, *Ethology* **113**, 821-824 (2007).

15    Tuci E., Ampatzis, C., Vicentini, F. & Doring, M. Evolved homogeneous neuro-controllers for robots with different sensory capabilities: coordinated motion and cooperation. *Proc. 9th international conference on From Animals to Animats: simulation of Adaptive Behavior* (*SAB'06*) (2006).

16    Buehler, J. Capabilities in Heterogeneous Multi-Robot Systems. *AAAI Conference on Artificial Intelligence* (2012).

17    Couzin, I. D., Frause, J., Franks, N. R. & Levin, S. A. Effective leadership and decision-making in animal groups on the move. *Nature* **433,** 513-515 (2005).

18    King, A. J., Johnson, D. D. P. & Van Vugt, M. The Origins and Evolution of Leadership. *Curr. Biol.* **19,** R911-R916 (2009).

19    Conradt, L. Models in animal collective decision-making: information uncertainty and conflicting preferences. *Interface Focus* **2**, 226-240 (2012).

20     Ward, A. J. W., Herbert-Read, J. E., Sumpter, D. J. T. & Krause, J. Fast and accurate decisions through collective vigilance in fish shoals. *PNAS* **108,** 2312-2315 (2011).

21    Reebs, S. G. Can a minority of informed leaders determine the foraging movements of a fish shoal? *Anim. Behav.* **59,** 403-409 (2000).

22    Couzin, I. D., Ioannou, C. C., Demirel, G., Gross, T., Torney, C. J., Hartnett, A., Conradt, L., Levin, S. A., & Leonard, N. E. Uninformed Individuals Promote Democratic Consensus in Animal Groups. *Science* **334,** 1578-1580 (2011).

23    Sueur, C. & Petit, O. Organization of group members at departure is driven by social structure in Macaca. *Int J. Primatol.* **29**, 1085-1098 (2008).

24    Šárová, R., Špinka, M., Panamá, J. L. A. & Šimeček, P. Graded leadership by dominant animals in a herd of female beef cattle on pasture. *Anim. Behav.* **79,** 1037-1045 (2010).

25    Guttal, V. & Couzin, I. D. Social interactions, information use, and the evolution of collective migration. *PNAS* **107**, 16172-16177 (2010).

26    Goldberg, D. E. *Genetic Algorithms in Search, Optimization, and Machine Learning* (Addison-Wesley, New Jersey, 1989).

27    Mones, E., Vicsek, L., & Vicsek, T. Hierarchy Measure for Complex Networks. *PLoS ONE* **7** e33799. doi:10.1371/journal.pone.0033799 (2012) 28    Harris, K. M. *The National Longitudinal Study of Adolescent Health (Add Health)* Waves I & II, 1994-1996; Wave III, 2001-2002; Wave IV, 2007-2009. (Chapel Hill, NC: Carolina Population Center, University of North Carolina at Chapel Hill, 2009).

29    Sueur, C., Deneubourg, J.-L. & Petit, O. From Social Network (Centralized vs. Decentralized) to Collective Decision-Making (Unshared vs. Shared Consensus). *PLoS ONE* **7(2)**: e32566. doi:10.1371/journal.pone.0032566.

30    Bindel, D., Kleinberg, J. & Oren, S. How Bad is Forming Your Own Opinion? *Proc. 52nd IEEE Symposium on Foundations of Computer Science* (2011).

31    Torney, C. J., Levin, S. A. & Couzin, I. D. Specialization and evolutionary branching within migratory populations. *PNAS* **107,** 20394–20399 (2010)

32    Johnstone, R. A. & Manica, A. Evolution of personality differences in leadership. *PNAS* **108,** 8373-8378 (2011)

33    Lorenz, K. *On Aggression* (University Paperbacks, San Diego, 1963)

34    Hemelrijk, C. Self-Organisation and Evolution of Biological and Social Systems (Cambridge University Press, July 21, 2011)

35    Dubreuil, B. Human Evolution and the Origins of Hierarchies: The State of Nature. (Cambridge University Press, Cambridge, 2010)

36    Bonabeau, E., Theraulaz, G. & Deneubourg, J.-L. Phase diagram of a model of self-organizing hierarchies. *Physica A,* **217,** 373-392 (1995)

37    Bonabeau, E., Theraulaz, G. & Deneubourg, J.-L. Mathematical model of self-organizing hierarchies in animal societies. *Bull. Math. Biol.* **58,** 661-717 (1996)

38    Hemelrijk, C. K., An individual-orientated model of the emergence of despotic and egalitartian societies. *Proc. R. Soc. Lond. B.* **266,** 361-369 (1999)

39    Barnard, C. J. & Silby, R. M., Producers and scroungers: a general model and its application to captive flocks of house sparrows. *Anim. Behav*. **29,** 543-550 (1981)

40    Giraldeau, L. A. & Caraco, T. *Social Foraging Theory* (Princeton University Press, Princeton NJ, 2000)

41    Tania, N., Vanderlei, B., Heath, J. P. & Edelstein-Keshet, L. Role of social interactions in dynamic patterns of resource patches and forager aggregation. *PNAS*, **109,** 11228-33 (2012) 42      Hamilton, B. H., Nickerson, J. A. & Owan, H. Team Incentives and Worker Heterogeneity: An Empirical Analysis of the Impact of Teams on Productivity and Participation. *J. Pol. Econ.* **111**, 465-497 (2003).

**Acknowledgements**

This research was partially supported by the EU FP7, ERC COLLMOT project No:227878

**Author Contribution**

A.Z. and T.V. designed the models. A.Z wrote and ran the programs. A.Z. and T.V. evaluated the data. T.V. wrote the manuscript.

**Competing Financial Interests**

The authors declare no competing financial interests.

# Supplementary Information

# Group performance is maximized by hierarchical competence distribution

Anna Zafeiris & Tamás Vicsek

**Supplementary Figures**

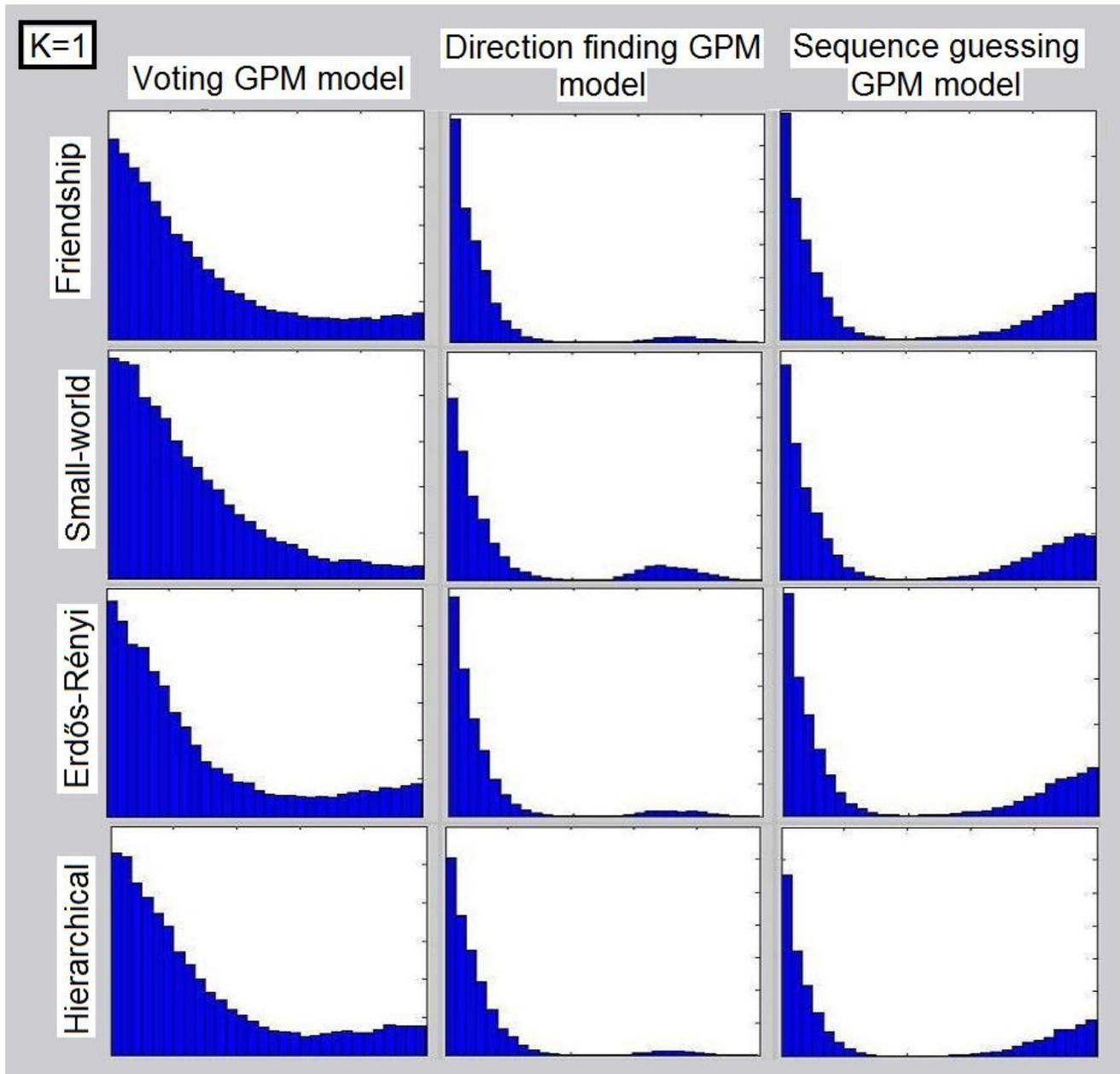

**Supplementary Figure S1**

**The optimal competence distributions for all network types and Group Performance Maximization models.** Group Performance Maximization models (GPMMs) for *K*=1, where *K* is the "cost of learning". This is the distribution that ensures the highest group performance with limited average competence. These are always highly skewed functions often with a structured tail.

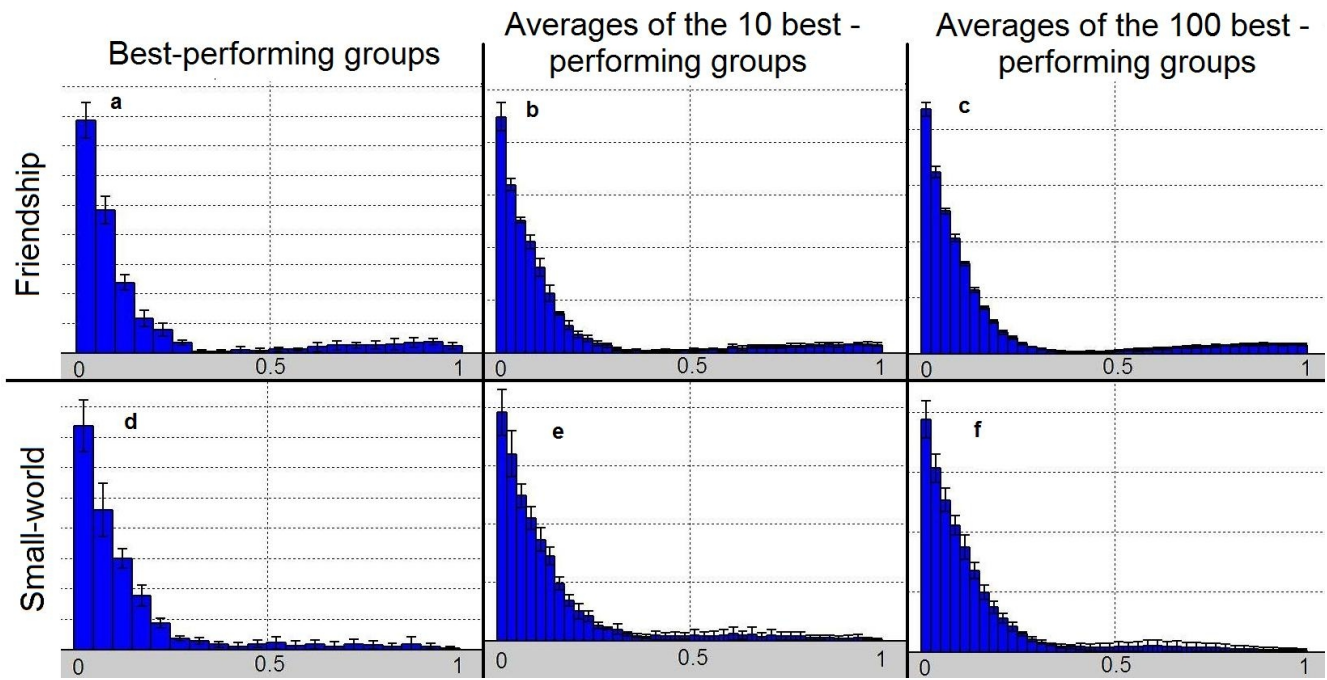

**Supplementary Figure S2**

**Standard deviation of 5 independent optimizations.** The results depicted in the first row, *a-c*, belong to the "Friendship" network, the second row, *d-f*, belong to 5 different small-world networks. In contrast with the Friendship network, which is a well-defined network determined by an extensive study, the small-world graphs are generated separately before each run of optimization. This calls forth the bigger error bars in the second row.

Sub-figures *a* and *d* depict the standard deviation of the 5 best performing groups belonging to the 5 independent optimizations. After this, we have determined the average distributions of the 10 and 100 best performing groups, for all five runs, respectively. The second column, *b* and *e* shows the standard deviation of the averages of the 10, *c* and *f* of the 100 best-performing groups, respectively.

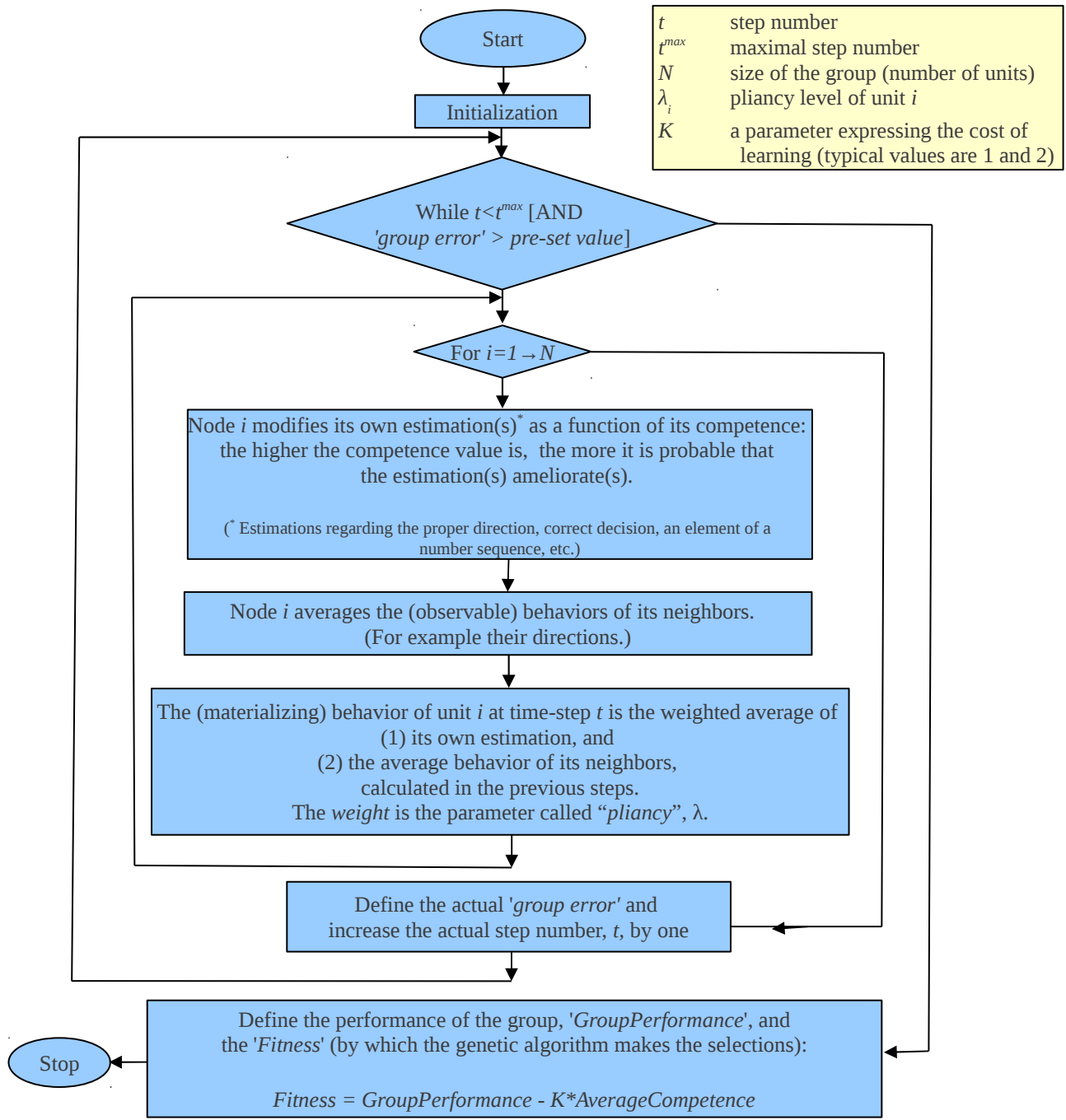

**Supplementary Figure S3**

**The general sketch of the GPMMs.** Since each unit had different competence level, the accuracy of their performance differed considerably (first rectangle in the for-loop). However, they could "learn'" from each other by observing (and "averaging'") the behaviour of their neighbours (second rectangle in the for-loop). The final, "materialized" behaviour of unit $i$ at time step $t$ was the weighted average of (i) its own estimation and (ii) of the average behaviour of its neighbours. The weighting factor, $\lambda_i$, also differed for each element. This parameter, "pliancy" expressed the willingness of the given unit to comply with its group mates. (The first and second rectangles in the for-loop are interchangeable.)

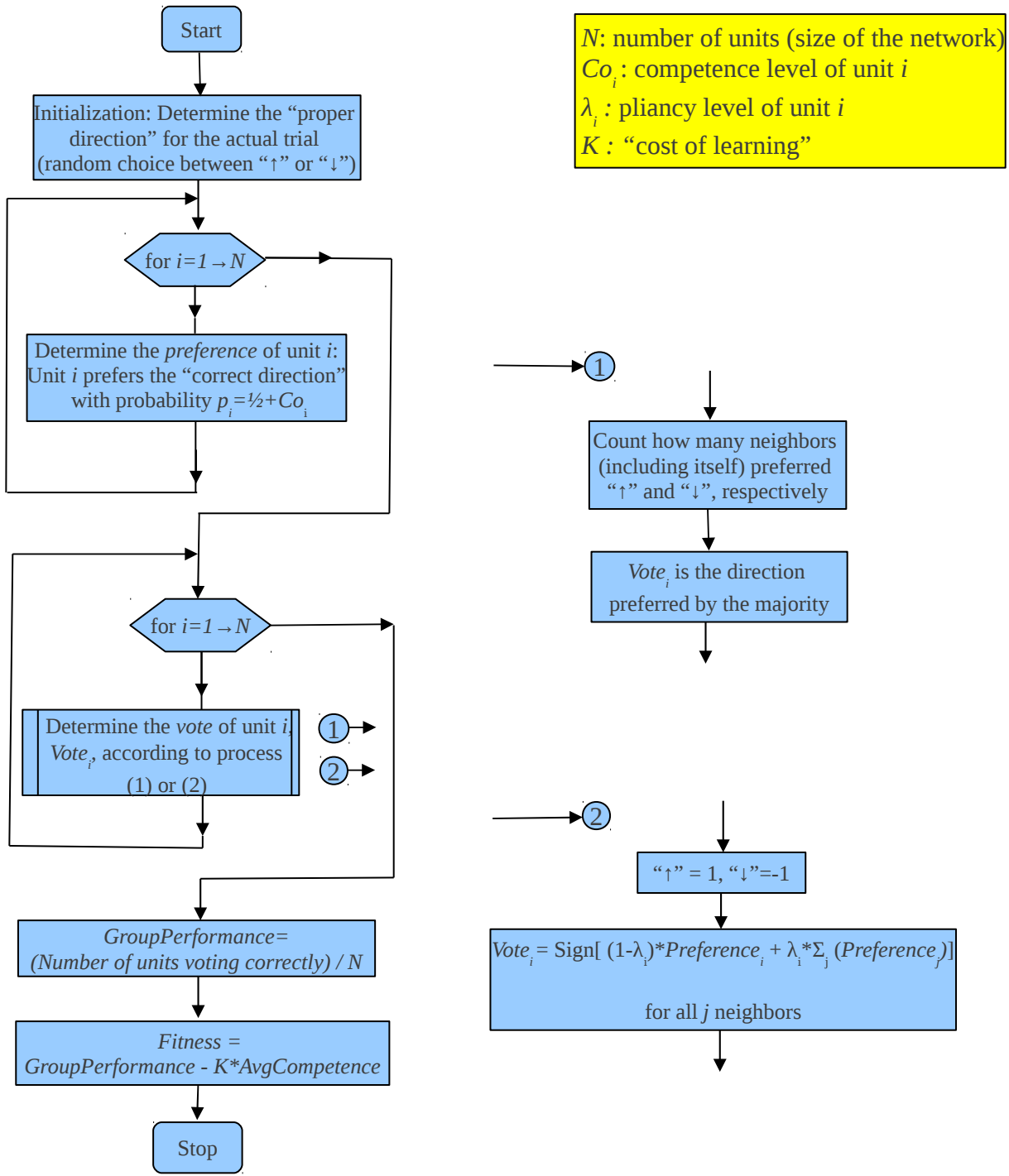

**Supplementary Figure S4**

**The flowchart of the "Voting GPMM".** In order to keep this model as simple as possible, we restricted it to two steps: (i) making up the preferences, and (ii) deciding on the final votes. In one version of the GPMM the units do not differentiate between their own preferences and that of their neighbours (they are weighted equally, marked with '1' on the figure), while in the other version a pliancy parameter is weighting the preferences (marked with '2').

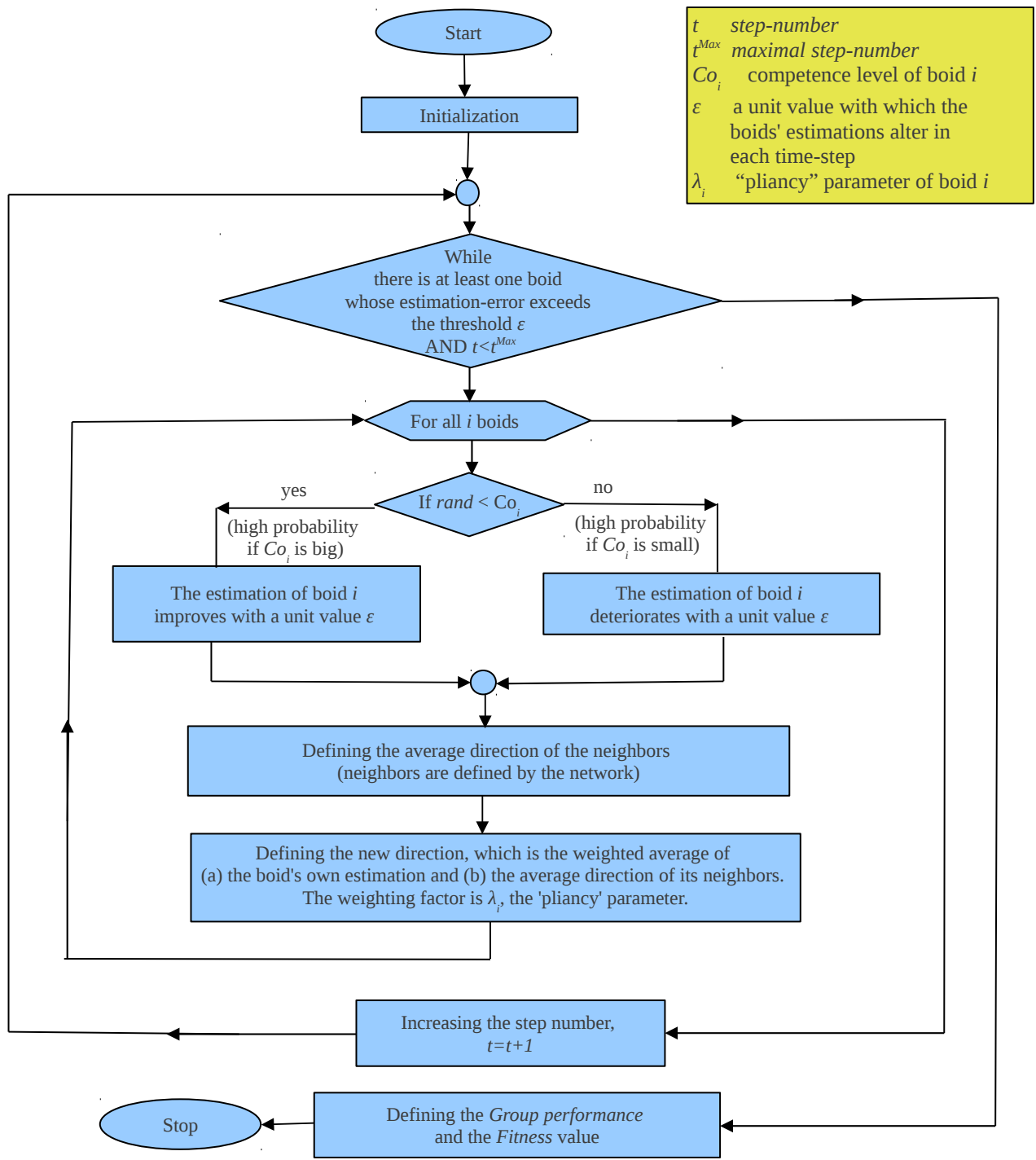

**Supplementary Figure S5**

**Flowchart of the "Direction Finding GPMM" in which the units have to find a pre-defined 'proper' direction.** Each unit at each time step modifies its direction-estimation with a unit angle ε, according to its competence level $Co_i$. The adopted direction in the given time step will depend partly on this estimation, and partly on the direction of the neighbours in the previous step.
The group performance $Pe$ is defined as

$$Pe_{DirectionFinding} = \sum \left( AvgDirErr_{t-1} - AngDirErr_t \right) / \pi = \left( AvgDirErr_0 - AvgDirErr_{t^{Max}} \right) / \pi$$ ,where

'$AvgDirErr$' is the average direction error, interpreted in radian.

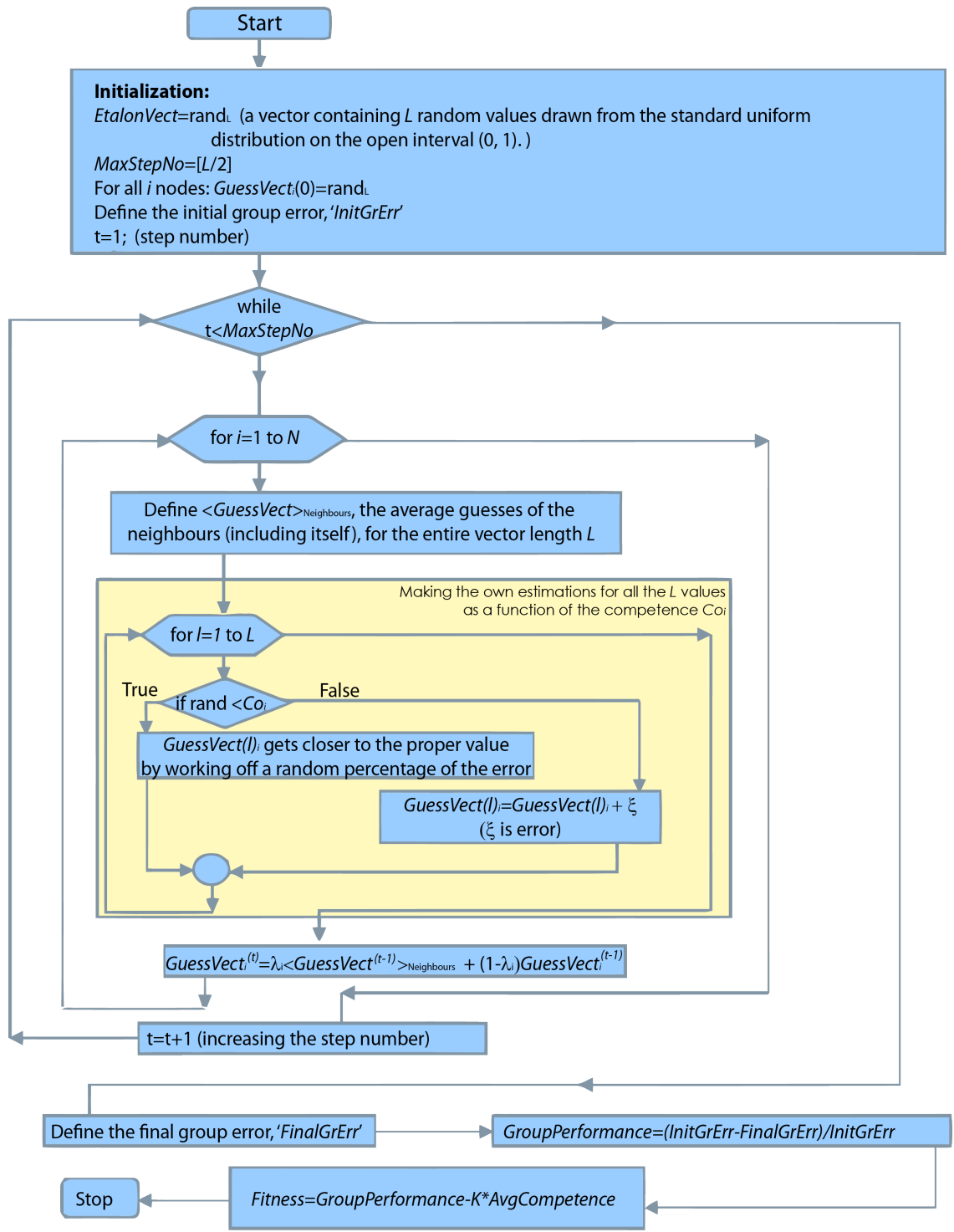

**Supplementary Figure S6**

**Flowchart of the "Sequence guessing GPMM".** Given a pre-defined number sequence and its random initial estimations for each unit. The aim is to reduce the initial estimation error as much as possible within a given number of steps. The group performance defines the ratio of the initial error that has been worked off:

$$Pe_{SequenceGuessing} = \frac{InitialGroupError - FinalGroupError}{InitialGroupError}$$

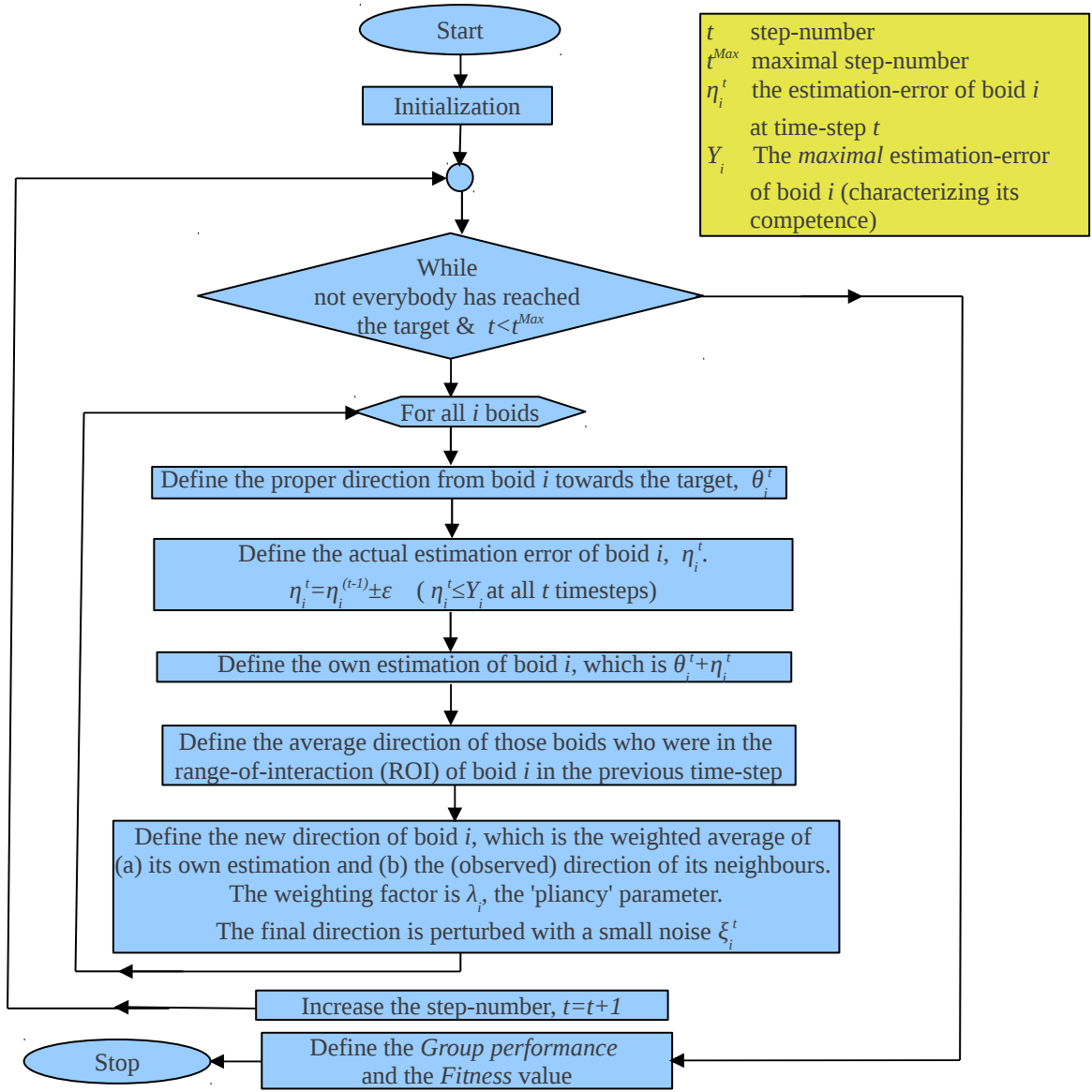

**Supplementary Figure S7**

**Flowchart of the "Flocking GPMM".** In this - somewhat separate- task the units were not motionless, organized into a graph, but were moving on a 2D surface aiming to reach a pre-defined target. The original motivation of this model was to model the homing flights of flocks of homing pigeons.

Here the group performance $Pe$ depended on three values: (i) the speed (number of steps) by which the flock reached the target, (ii) if the units were moving (navigating) together or not, and (iii) how many boids got lost. Precisely:

$$Pe_{Flocking} = \frac{D}{\langle s_i \rangle_T} \cdot \left(1 - \left\langle \frac{s_i^{out}}{s_i} \right\rangle_T \right) \cdot \left(1 - \frac{N_{Lost}}{N}\right)$$, where $D$ is the distance between the starting point and the

target, $s_i$ is the number of steps unit $i$ needed to reach the target, $s_i^{out}$ is the number of steps in which unit $i$ was "alone" (that is, its ROI was empty), $N_{Lost}$: number of lost units, $N$: number of all units ("flock size"), $<...>$ averaging (over all nodes of the graph), and $<...>_T$ is the averaging over those elements which reached the target.

**Individual vs. group optimization: a comparison of the optimal competence distributions**

In order to identify the differences in the optimal competence distributions in case of individual and group optimizations, we have carried out experiments on various group sizes using the flocking model. Apart from the fitness functions, all the parameters and settings were the same. The fitness function in the case of group optimization was the same as the one we used in order to perform the optimizations reported in the present paper, and in case of individual optimization it was defined for individual *i as follows*:

$F_i=D/s_i-K\cdot Co_i$, where $D$ is the distance between the starting point and the target, $s_i$ is the number of steps unit $i$ needed to reach it, $K$ is the '*cost-of-learning*' and $Co_i$ is the competence value of the focal individual. (For comparison see also the subscription of Supplementary Figure 7.)

The corresponding results are depicted below. As it can be seen, individual optimization results in two much more sharply separated subgroups: individuals are either "informed" ones (with relative high competence values) or they belong to the uninformed subgroup in which the competence values are zero or almost-zero. Units with in-between competence levels do not appear, which is related with the more sharp slam of the very small competence values that can be observed in the upper row. In contrast, group optimization favours the appearance of individuals with more diverse characteristics. As the size of the flock is increasing, these differences are getting more and more pronounced.

It is also important to note that the two kinds of distributions do not work against each other in the sense that they do not act in a way that one would lessen the other. Accordingly, in case these two effects (group and individual selection) appear together, the result is - most probably - a distribution somewhere in between the two "clear" cases.

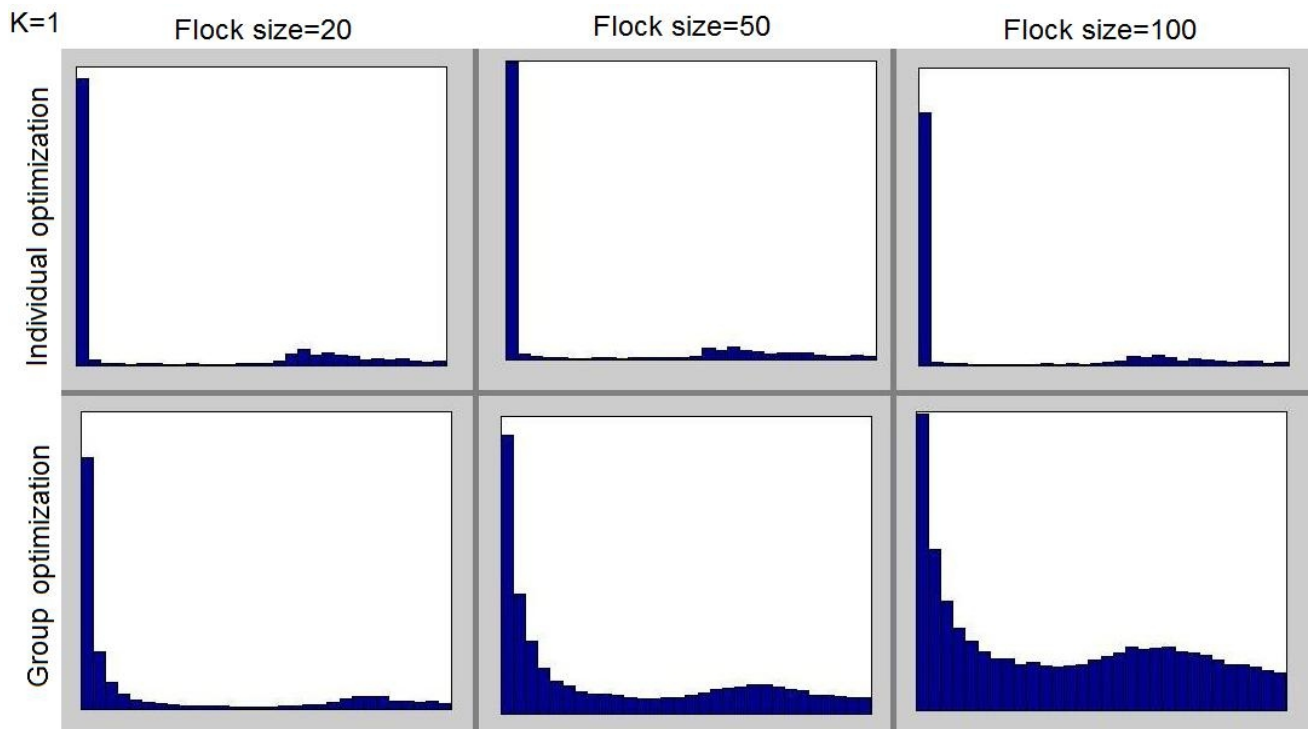

**Supplementary Figure S8**

Optimal competence distributions for the flocking model in case of individual (upper row) and group (bottom row) optimization, for three different group sizes: N=20 (1[st] column), 50 (2[nd] col.) and 100 (3[rd] col).

**Supplementary Methods**

**Detailed description of the networks**

In three out of the four Group Performance Maximization models (GPM models or GPMMs), the units were motionless and the interaction among them was defined by the structure of a network. In these models (Voting, Sequence guessing and Direction finding GPMMs) two nodes exchanged information if there was an edge between them and did not communicate otherwise. We examined the group performance on four different network types, all of which were static and connected with $N=200$ nodes, except for the friendship network which contained $N=204$ vertices.

**1. Friendship network**

This real-world network describes the friendship relations of adolescents in grades 7-12, collected by "The National Longitudinal Study of Adolescent Health (Add Health)"' in the United States during the 1994-95 school year[28]. In this study, 80 high schools and 52 middle schools were selected with unequal probability in order to ensure the sample to be representative with respect to region of country, urbanicity, school size, school type, and ethnicity. Out of this data set we selected community no. 8 because of its size: it contains 205 nodes (representing the friendship relations of 205 students). Students did not necessarily name each other as friend symmetrically, so the resulting graph is directed. One of the nodes were not connected to the rest of the graph (nobody claimed that he/she is a friend of her/him, and vice versa, this student said she/he has no friends). This vertex was not considered, so the GPMMs run on a network containing $N=204$ nodes. (Figure. 1a in the main article shows the undirected version of this graph.)

**2. Small-world networks**

This network type was generated by us in the following way:

a) Firstly we connected each node to their first and second neighbours, then

b) we randomly selected $p^{SW}$ ratio of the edges,

c) the selected links were deleted and each original start-node was connected to another, randomly selected end-node

In our case, the parameters were $N=200$ and $p^{SW}=0.15$, and the resulting graph is undirected.

**3. Erdős-Rényi graphs**

In order to generate an Erdős-Rényi network, we considered all the possible vertex-pairs and created an edge between them with the probability $p^{ER}=0.015$. Note, that the percolation threshold for an Erdős-Rényi graph is $p^{ERpercol} = 1/N$, that is, for $N=200$, $p^{ERpercol} = 1/200=0.005$, which is one third of the parameter we have chosen, $p^{ER}=0.015$. Accordingly, our graph had a high chance to consist of one giant component, but in order to make sure that the created network is connected, a subroutine determined all the occurrent components and disconnected nodes, and connected them to the giant component. The resulting network is undirected.

## 4.  Hierarchical network

This directed graph was generated according to the method described in[27]. This process starts from a tree-structure and then adds further edges randomly, according to a pre-defined hierarchy parameter. We used a graph characterized by the parameter $p^{Hier}=0.8$.

Importantly - and somewhat counter intuitively - the structure of the network did *not* have a fundamental effect either on the distribution of the competence values within the optimized groups (see Supplementary Figure S1 and Figure 2 in the main article), nor on the group performance *Pe*. By varying the type of the network and keeping all other parameters unchanged, the difference between the *Pe* values is less than 5 percent (in the sense that $Pe^{Max}$-$Pe^{Min}$ (≈ 0.69-0.65) ≤ 0.05, which is 5% of the [0,1] interval from where *Pe* can take values. This remains true in spite of the fact that there are big differences between the edge numbers of the various graphs that we used: as mentioned in the Methods Summary, $E^{Frnd}$=1012, $E^{Hier}$=777, $E^{ER}$ ≈ 310 and $E^{SW}$ ≈ 400. In case of the hierarchical and friendship networks these numbers are given, since the graphs themselves are given. But by varying the Erdős-Rényi and small-world networks in a way that $E^{ER} \approx E^{SW} \approx 800$ and executing the experiment again, we get that the above mentioned difference, $Pe^{Max}$-$Pe^{Min}$ is approximately 1 percent, that is, even smaller. These test were made by using the sequence guessing GPMM.

**Commonalities among the models:**

Each model had to satisfy the following criteria:

- The group, as a whole, had to perform a task, (like finding a location, direction, estimating a number sequence or finding out a 'proper' decision by voting).
- The group performance – characterizing the efficiency of the collective decision making process – was measurable and quantifiable, typically taking values from the [0...1] interval. Better performance corresponded to values closer to 1.
- Each unit contributed to finding the best solution with varying degrees of input, depending on its competence level.
- The average (or total) competence was commensurable and quantifiable as well. (Here the two expressions, "average" and "total" are equivalent, since the number of units, $N$ was fixed at the beginning of each trial).
- Only the neighbouring units could communicate with each other (that is, exchange information, copy each others behaviour, etc.). In three models out of the four, the communication structure was defined by graphs: two units could exchange information, if there was an edge between them. In the fourth case, which we called as the Flocking model the actors were moving, and thus the graph of interactions was changing in time: those particles interacted with each other which were closer than a pre-defined distance called "Range of Interaction, ROI".

**Differences among the models**

- In one of the models (Flocking) the units move and the graph of interactions is time dependent, while in the other three models static graphs were assumed to specify the structure of interactions.

- We considered four different models: a) choosing the true solution out of two options, b) giving a good estimate of the values of a series of numbers, c) guessing a particular direction and d) finding a position on a plane. Correspondingly, the definitions of the group performance were different – but following the same general pattern.

- The definition of competence is also model dependent. In b) and c) competence was proportional to the probability that the unit at the given time step improves its estimate, while in the case of d) it was related to the maximal error in the estimation of a unit.

- The initial conditions were taken as random values (within a given interval) except model a) where the initial state of a unit was already correct in proportion of its competence.

- If the competence level of a unit was so low that it had no chance to improve its estimation in the given step, than we either changed its estimation by a small random value (b), we have made its estimation less accurate (c) or changed the value of the estimation by a given small number randomly (up or down).

We had a particular reason to introduce differences listed above: our goal was to demonstrate that the main result of ours, i.e., that the multiple hierarchy solution is the optimal one in most of the possible situations is not dependent on the various details of the models.